\pdfoutput=1
\documentclass[11pt]{amsart}
\usepackage{tikz,blkarray,multirow,commath}
\usepackage{graphicx,euscript,verbatim,constants}
\usepackage{amssymb,url,booktabs}
\usepackage{todonotes}

\newtheorem{Thm}{Theorem}
\newtheorem{Cor}[Thm]{Corollary}

\newtheorem{lemma}[Thm]{Lemma}

\newcommand{\nc}{\newcommand}
\nc{\eu}{\EuScript}
\nc{\on}{\operatorname}
\nc{\M}{\eu{M}}
\newconstantfamily{const}{symbol=c}

\nc{\utau}{\underline{\tau}}
\nc{\otau}{\bar{\tau}}
\nc{\urho}{\underline{\rho}}
\nc{\uj}{j'}
\nc{\diag}{\on{diag}}
\nc{\bX}{\mathbf{X}}
\nc{\bA}{\mathbf{A}}
\nc{\mE}{\mathcal{E}}
\nc{\tmu}{\tilde{\mu}}
\nc{\tx}{\tilde{\xi}}
\nc{\ttau}{\tau}
\nc{\tsig}{\tilde{\sigma}}
\nc{\hA}{\hat{A}}
\nc{\hR}{\hat{R}}
\nc{\ha}{\hat{a}}
\nc{\hD}{\hat{D}^{*}}

\nc{\constc}{c}
\nc{\constcp}{c'}
\nc{\J}{\eu{J}}
\nc{\wt}{\widetilde}
\nc{\tB}{\wt{B}}
\nc{\tf}{\tilde{f}}
\nc{\tD}{\wt{\Delta}}
\nc{\U}{\mathcal{U}}
\nc{\V}{\mathcal{V}}
\nc{\tU}{\check{U}}
\nc{\tV}{\check{V}}
\nc{\tN}{\wt{N}}
\nc{\tY}{\wt{Y}}
\nc{\tX}{\wt{X}}
\nc{\E}{\mathbb{E}}
\nc{\R}{\mathbb{R}}
\nc{\Les}{\mathcal{L}}
\renewcommand{\P}{\mathbb{P}}
\nc{\e}{\mathrm{e}}
\nc{\var}{\on{Var}}
\nc{\F}{\eu{F}}
\nc{\diam}{\on{diam}}

\nc{\indic}{\mathbf{1}}
	
\begin{document}
\title[Stochastic growth rates with rare diapause]{Stability of fixed life histories to perturbation by rare diapause}
\author{David Steinsaltz} 
\address{David Steinsaltz\\Department of Statistics\\University of Oxford\\24--29 St Giles\\Oxford OX1 2HB\\United Kingdom}
\email{david.steinsaltz@stats.ox.ac.uk}
\author{Shripad Tuljapurkar}
 \address{Shripad Tuljapurkar\\454 Herrin Labs\\Department of Biology\\Stanford University\\Stanford CA 94305-5020\\USA}
 

\begin{abstract}
	We analyze the behavior of an age-structured population subject to stochastically
	varying linear survival and reproduction at age-dependent rates, in the special case
	where births occur only when organisms attain a fixed maximum age $d$, so that
	generations have a constant length $d$.
	We show that perturbing this fixed-length life history by a small diapause --- a
	delay in development, corresponding to adding diagonal terms of size $\epsilon$ to
	the matrix that updates the population vector from one time period to the next ---
	increases the asymptotic stochastic growth rate by an increment of order
	$(\log\epsilon^{-1})^{-1}$, and at least $\frac{\sigma_*^2}{2\pi d\log\epsilon^{-1}}$, where
	$\sigma_*^2$ is a sum of variances of log ratios of survival and birth rates one age
	class apart.
	The growth rate is thus continuous but not differentiable at $\epsilon=0$.
	As this effect dominates any linear cost suffered by individuals who are subject to
	diapause, it follows that a small random disruption to the deterministic life history
	would be favored by natural selection, in the sense that it would increase the
	stochastic growth rate relative to the zero-delay deterministic life history.
	We prove this in the wider setting of matrix migration models in which two or more
	sites share the maximum mean growth rate, a degeneracy that is excluded in the treatment of
	\cite{migration2018,jaggi2023migration}, where the growth rate instead increases as a
	power of $\epsilon$.
\end{abstract}
\maketitle

\section{Introduction}
\subsection{Biological motivation}
In considering the evolution of developmental delays, it is crucial to consider the effect on population fitness of perturbations around a base state where organisms
are constrained to a fixed developmental sequence.
It has long been argued \cite{cole1954pcl} that populations of individuals who delay or spread reproduction over time will suffer reduced growth rate. Within the framework of matrix population models in a deterministic environment --- where demographic rates are the same every year --- this follows from a theorem of Karlin \cite{karlin82}.

Cohen \cite{cohen1966} and Cohen and Levin \cite{cohen1991}, on the other hand, used analysis and simulations to show that long-run growth of a population could benefit from a life cycle delay when there are some kinds of random variation in time, or by migration when there are some kinds of random variation across space. 
These kinds of stochastic variation have been formulated as random matrix models, whose Lyapunov exponent is the long-run growth rate of the population, as discussed by \cite{sT90delayed,tuljapurkar2000ets, Wiener1994}. In this general setting, we would like to know whether the long-run growth rate increases when there is mixing in time \cite{tuljapurkar2000ets} --- biologically, when should delay be favored to evolve? 
A general and precise answer has been difficult because previous work \cite{Wiener1994} shows that the long-run growth rate can be singular (e.g., non-differentiable) in the limit of no mixing, hence inaccessible to standard methods that depend on calculating or approximating a derivative in the diapause (or migration) rate. 
A similar singularity arises in random-matrix models used in models of disordered matter \cite{derrida1983singular}. 

We note that the genetic consequences of populations experiencing diapause and
dormancy have been the subject of considerable mathematical interest
\cite{shoemaker2018seedbank,blath2016seedbank,heinrich2018effects,lennon2021principles};
and the question of when dormancy is favored has lately been approached from that
direction as well. Etheridge and de Oliveira Madeira \cite{EM26} show, for a
Wright--Fisher model with seed bank, that the establishment of a dormancy-producing
mutant turns on the trend in population size, dormancy being favored in a declining
population. The effect we describe here is of a different kind, driven by the
variance of the environmental fluctuations rather than by any systematic trend, and
measured by the stochastic growth rate rather than by establishment under drift.
Closer to the present work, Blath, Hermann and Slowik \cite{BHS21} compare dormancy
strategies in a two-type branching process in a randomly fluctuating environment,
taking the maximal Lyapunov exponent as the measure of fitness, and obtain a map of
the environmental regimes in which each strategy is favored; a survey of this line
of work may be found in \cite{blath2024dormancy}.

The growth-rate effects of diapause in stochastic environments were analyzed for
special cases in \cite{TI93}. There, as in \cite{BHS21}, the cases that yield exact
expressions are essentially those in which the system reduces to a one-dimensional
problem, and the methods applied are unable to shed light on the behavior near the
crucial boundary of zero diapause.
There has also been application of simulation methods to these problems, such as
\cite{EE00,menu2000bet}, but simulation approaches that boundary only slowly: the
simulations of \cite{menu2000bet}, for instance, use $0.1$ as the smallest diapause
rate.
In the analogous but fundamentally different diapause model of \cite{magalie2023optimal} --- active lifespans there have indeterminate length, so the degeneracy which concerns us here does not arise --- the finest resolution of  diapause rates considered in simulations is still only $0.025$.
Current computer technology makes it easy to explore far deeper into the
singularity, as we do in section \ref{sec:simulation}, corroborating our theoretical
results; but simulation alone casts little light on the underlying reasons why even
extremely rare diapause may still be substantially favored.
As far as we are aware, the present paper gives the first analytic description of the
stochastic growth rate in a neighborhood of zero diapause --- an earlier version of these results was posted as \cite{diapause2018} in 2018 --- and hence the first proof
that a fixed deterministic life history is evolutionarily unstable with respect to
perturbation by diapause.
The work also fits comfortably into the conceptual framework of ``bet-hedging'' in
stochastic environments, as described by \cite{menu2000bet}, while providing more
detail about the nature of the realized ancestries that end up comprising the rare
``winning bets''.

What we show, in Theorem \ref{T:samerate1}, is that in a basic matrix model of diapause under very general conditions
the sensitivity of stochastic growth rate to changes in
diapause rate $\epsilon$ varies near 0 like $1/\log \epsilon^{-1}$. 
This implies that a sufficiently small delay will always
increase the population growth rate. 
It is a reasonable interpretation that the semelparous life history is evolutionarily unstable, and that
some diapause will be favored by natural selection, regardless of the cost due to
increased mortality or lost reproduction among those
suffering the delay.

This behavior may be contrasted with that found by \cite{HFRM26} for a delayed-logistic model in which every individual matures after a fixed lag, in an environment fluctuating with finite correlation time; there the growth rate is maximal at zero lag, and environmental variability strictly increases the cost of delay. In that setting all individuals share a single fluctuating environment, so a delay merely re-times access to it. 
The mechanism studied here is of a different character: the semelparous life history partitions the population into lineages that accumulate distinct environmental histories, and it is the rarity of the diapause --- allowing an ancestry to shift between those lineages at opportune moments --- that produces the singular gain.

\subsection{Connection to migration models}
Here we consider a random-matrix model of life histories represented by ``random Leslie matrices'', as pioneered in \cite{sT90}.
Each time period is associated with a $d\times d$ matrix $\Les_t$, whose subdiagonal elements $S_t^{(j)}$ represent fractions of the population aged $j$ surviving to the next time period --- which we will refer to, for convenience, as ``years'' --- and whose top row $(B_t^{(j)})$ represents the number of offspring in age class 0 produced by each individual in age class $j$.
We focus here on the ``semelparous'' setting created by having all birth rates $B_t^{(j)}=0$ except the last, in age class $j=d-1$.
This life history is singular, in that the population decomposes into $d$ nonoverlapping lineages, comprising individuals separated by exactly $d$ years.
We then proceed to introduce a small perturbation, adding diagonal elements that represent diapause, proportional to a tunable quantity $\epsilon$, that breaks this decomposition.
The goal is to analyze the mathematical singularity in the long-term population growth rate, the top Lyapunov exponent, induced by a perturbation of size $\epsilon$ in a neighborhood of 0.

There is a strong analogy between this work and the previous work \cite{migration2018,jaggi2023migration}, which analyzed the behavior of a multisite population model where a small amount of migration was introduced into an otherwise sessile population.
Indeed, this is very close to an exact analogy if we look at the population at intervals of $d$ years, so considering the product of $d$ Leslie matrices in a bunch.
If we translate ``sites'' to ``age classes'', we see that the semelparous life history becomes a ``non-migratory'' structure, where individuals of age class $j$ produce a succeeding generation of some random size, all in exactly the same age class, $d$ years later.
Introducing diapause at rate proportional to $\epsilon$ is then equivalent to introducing migration from class $j$ to $j-1$ in the $d$-year model, also at rate proportional to $\epsilon$.
The one formal difference is that there will also be additional terms further of the diagonal proportional to higher powers of $\epsilon$; it is easy to see that these higher-order terms have no significant effect on the asymptotics for $\epsilon \to 0$.

The diapause model is not a special case of the previously analyzed migration models for one essential reason:
A key condition in \cite{migration2018} and \cite{jaggi2023migration} was that there is a single site that has the maximum average growth rate.
The main results state that the top Lyapunov exponent grows, in a neighborhood of $\epsilon=0$, like a power $\epsilon^\rho$, where the power $\rho$ would vanish when the difference between the best and second-best sites goes to 0.
When the diapause model is transformed into a migration model, it is inevitable that all sites have exactly the same average growth rate, namely
\begin{equation}
	\label{E:averagegrowth}
	\E\bigl[\log B \bigr] + \sum_{j=0}^{d-2} \E\bigl[\log S^{(j)} \bigr].
\end{equation}
This fundamentally changes the asymptotic behavior, and also requires substantially different methods of analysis.
We will thus subsume the diapause model into a more general model: Matrix migration models in which two or more sites have equal average stochastic growth rates.

\section{Formal definition of the model and results}
\subsection{Diapause life history}  \label{sec:diapause}
Consider a population in which individuals progress through immature life stages until reaching adulthood, when they reproduce and
then die. 
Diapause is a life-cycle delay in which individuals can stay in some immature stage with some probability. We can describe diapause
by reconceptualizing the ``sites'' of the previous section as life stages, and also describe an organism's progress using matrices that are not diagonal, but sub-diagonal. 
The life stages  (or sites) are viewed as a cycle, described by Leslie matrices of the form
$$
\Les_{t}:= 
\begin{pmatrix}
	0&0&\cdots&0&B_{t}\\
	S_{t}^{(0)}&0&\cdots&0&0\\
	0&S_{t}^{(1)}&\cdots&0&0\\
	\vdots&\vdots&\ddots&\vdots&\vdots\\
	0&0&\cdots&S_{t}^{(d-2)}&0
\end{pmatrix}
$$
Here ages run from $1$ to $d$, and are equivalently referred to as age classes that run from 0 to $d-1$. The quantity $S_{t}^{(j)}\in (0,1)$ is the proportion surviving from age $j$ to $j+1$ in year $t$, and $B_{t}$ is the average number of offspring produced when an individual becomes mature in age-class $d-1$.
Offspring are born into age-class 0, and the parent --- in age class $d-1$ --- dies.
To this we add a random diagonal entry $\epsilon \widehat{A}^{(j)}$, corresponding to a small fraction of the population aged $j$ that remains for one year at age $j$, rather than advancing.

Considering now the $d$-step population process $N_{sd}$ (which may be understood as looking at the population age distribution in timesteps corresponding to a full $d$-year generation), the matrix that transforms $N_{sd}$ to $N_{(s+1)d}$ is
\begin{equation}
	\label{E:diapauseDs}
	 \begin{pmatrix}
		\e^{X_s^{(0)}} & \epsilon A_s^{(0)} & 0&\cdots & 0 & 0\\
		0 & \e^{X_s^{(1)}} & \epsilon A_s^{(1)} &\cdots & 0 & 0\\
		\vdots&\vdots&\vdots&\ddots & \vdots& \vdots\\
		0&0&0&\cdots & \e^{X_s^{(d-2)}}& \epsilon A_s^{(d-2)}\\
		\epsilon A_s^{(d-1)} & 0 & 0 & \cdots & 0 & \e^{X_s^{(d-1)}}
	\end{pmatrix}
	+
	\epsilon^2 A'_s,
\end{equation}
where
$$
X_s^{(j)} = \sum_{i=0}^{d-j-2} \log S_{sd+i}^{(i+j)} +  \log B_{sd+d-j-1}  + \sum_{i=d-j}^{d-1} \log S_{sd+i}^{(i+j-d)}.
$$
The off-diagonal terms $A_s^{(j)}$ are products of the diagonal diapause rates $\widehat{A}$, with survival rates and birth rates from the matrices $\Les_{sd},\Les_{sd+1},\dots,\Les_{sd+d-1}$, hence will be controlled by the same assumptions we impose on those parameters.
The same is true of the second-order error matrix $A'_s$.
For this full-generation population growth process we immediately have the mean baseline log growth rate given in \eqref{E:averagegrowth}, hence the asymptotic growth rate of the diapause model considered on the natural annual scale is
\begin{equation} \label{E:a0diapause}
	a(0)=\frac{1}{d}\Bigl(\E[\log B_{t}]+\sum_{j=0}^{d-2} \E[\log S_{t}^{(j)}]\Bigr).
\end{equation}
Consequently, we may apply Theorem \ref{T:samerate1} to this $D_{t}$, producing the same
$1/\log\epsilon^{-1}$ rate of increase, as stated in Corollary \ref{C:diapause}. 
The populations at different ``sites'' now correspond to subsets of the population in different age classes.

\subsection{The migration model} \label{sec:migration}
The mathematical setting is similar to that of \cite{migration2018}, though some of the particular
assumptions differ.
We suppose that populations $0, \dots, d-1$ have log growth
rates $X_t^{(i)}$ during time period $t$ at site $i$. The
vectors $X_t=(X_t^{(0)},\dots, X_t^{(d-1)})$ are assumed
independent, but the components within each $X_t$ ---
corresponding to growth rates at different sites at the same time
--- may be dependent. We assume these have finite expectation and variance,
writing ${\mu_i:= \E[X_t^{(i)}]}$, and $\mu:=\max_i \mu_i$. 
We write $\bX$ for the complete collection of all $X^{(j)}_{t}$ for ${j=0,1,\dots,d-1}$,
$0\le t<\infty$, and similarly define $\bA$.

We define the {\em migration graph} $\M$ to be a simple and irreducible directed graph 
whose vertices are the sites $\{0,\dots,d-1\}$, representing the transitions that have nonzero
probability.
We let $A_{t}$ be an i.i.d.\ sequence of nonnegative
$d\times d$ matrices with zeros on the diagonal, representing migration rates in time-interval $t$. 
We follow the convention from the matrix population model literature, that transition
rates from state $i$ to state $j$ are found in matrix entry $(j,i)$. 
Population distributions
are thus naturally column vectors $(N_t(j))_{j=0}^{d-1}$, and the updating from time $t-1$ to time $t$ is effected
by left multiplication.

We assume that
the collection of pairs $(X_{t},A_{t})_{t=0}^\infty$ is jointly independent, but note that 
we do not assume for a given $t$ that $A_{t}$ and $X_{t}$ are independent, or that different matrix entries
corresponding to the same $t$ are independent.
We assume that there is an $A^*$ such that
\begin{equation} \label{E:Astar}
	\log A_{t}(j,i)  \le A^*\text{ almost surely for all $i$ and $j$};
\end{equation}
and that $(A_t)$ is {\em adapted} to $\M$, by which we mean that
\begin{equation}
	\label{E:Azero}
	A_t(j,i) = 0 \text{ almost surely if and only if } i\nrightarrow j.
\end{equation}
It follows that there exist $A_*$ and $p>0$ such that if $i\to j$ then ${\P(\log A_t(j,i)\ge -A_*)\ge p}$.

The basic objects we will be analyzing are the matrices
$$
	D_{t}(\epsilon):= \begin{pmatrix}
		\e^{X_t^{(0)}} & \epsilon \e^{X_t^{(0)}} A_t(0,1) &\cdots  & \epsilon \e^{X_t^{(0)}} A_t(0,d-1)\\
		\epsilon \e^{X_t^{(1)}} A_t(1,0) & \e^{X_t^{(1)}}  &\cdots & \epsilon \e^{X_t^{(1)}} A_t(1,d-1)\\
		\vdots&\vdots&\ddots & \vdots\\
		\epsilon \e^{X_t^{(d-1)}} A_t(d-1,0) & \epsilon \e^{X_t^{(d-1)}} A_t(d-1,1) &\cdots & \epsilon \e^{X_t^{(d-1)}}
	\end{pmatrix}.
$$
Note the choice to multiply each row by the corresponding growth rate $\e^{X_t^{(j)}}$ from the diagonal, which is equivalent to saying that migration occurs at the beginning of a time step, after which the individuals who have migrated partake of the growth rate at their new site.
It is easy to see that this has no effect on any of the conclusions of our results below, as it affects the growth rate only at a fraction of time steps that becomes infinitesimal in the limit.
We could just as well absorb the factors $\e^{X_t^{(j)}}$ into the transition rates $A_t(j,k)$ without affecting any of the assumptions of Theorem \ref{T:samerate1}.
The only reason for choosing to write it this way here is that that it slightly simplifies the notation that we will introduce in section \ref{sec:Trajectories}.
We will not need to specifically highlight the dependence of $D_t$ on $\epsilon$, so we suppress $\epsilon$ from the notation, and instead use the brackets as with $A_t$ to denote matrix entries; thus, $D_t(j,k)$ is the $j,k$ entry of $D_t(\epsilon)$.
We define the partial products
$$
R_{T}(\epsilon):= D_{T-1}\cdot D_{T-2}\cdot\cdots\cdot D_{0}.
$$

For fixed $\epsilon>0$ the i.i.d.\  sequence $D_{t}$
satisfies the conditions for the existence of a stochastic growth rate independent
of starting condition.\cite{jC79} 
That is, 
$$
a(\epsilon):=\lim_{T\to\infty} T^{-1}\log R_{T}(\epsilon)_{ij}
$$
are well defined deterministic quantities, in the sense that the limit exists
almost surely, is almost-surely constant, and is the same for any 
${0\le i,j\le d-1}$.

Of course, $R_{T}(0)$ has a different structure: 
The off-diagonal terms
are all 0, while on the diagonal, by the Strong Law of Large Numbers,
$$
\lim_{T\to\infty} T^{-1}\log R_{T}(0)_{ii}=\mu_{i}.
$$

For the upper bounds on growth rate (see Theorem \ref{T:samerate1}) we will be assuming that the
demeaned differences  $(X_t^{(i)}-\mu_i) - (X_t^{(j)}-\mu_j)$ are {\em sub-Gaussian}.
Definitions and important properties of sub-Gaussian random variables may be found in the Appendix.
All Gaussian random variables are also sub-Gaussian.

For any cycle $\gamma$ in $\M$ we define
$\tilde\gamma$ to be the sequence of sites obtained
by removing sites from $\gamma$ that do not have the
maximum mean log growth rate --- that is, sites $\gamma_i$ such that $\E[X_t^{(\gamma_i)}]<\mu$.
($\tilde\gamma$ will, in general, not be a path in $\M$.)
We define
$$
	\sigma_\gamma^2 :=  \frac{1}{|\gamma|}\sum_{i=0}^{|\tilde\gamma|-1}\var ( X_t^{(\tilde\gamma_i)}- X_t^{(\tilde\gamma_{i+1})}).
$$
Then
\begin{equation} \label{E:definesigma}
\sigma_*^2 := \max_\gamma \sigma_\gamma^2,
\end{equation}
where the maximum is taken over all cycles $\gamma$ in 
$\M$. 
Note that the cycle may pass through any sites, but the variance is counted only for those sites with
optimal mean log growth rate, while the denominator
counts all sites in the cycle. 
This effectively penalizes cycles that pass through nonoptimal sites, though these still need to be considered, as they may yield
the maximum $\sigma_\gamma^2$ by passing through
other sites of higher variance.  
We give an example of computing $\sigma_*^2$ in Figure \ref{F:sigma_star1}.

\begin{figure}[ht]
	\begin{center}
		\begin{minipage}[t]{0.2\textwidth}
			$$\hspace*{-5mm}\M$$
			\begin{tikzpicture}[scale=1, transform shape]
			\tikzstyle{every node} = [circle, fill=gray!30]
			\node (3) at (0, 0) {3};
			\node (2) at +(2,0) {2};
			\node (0) at +(0,2) {0};
			\node (1) at +(2,2) {1}; 
			\foreach \from/\to in {0/2, 2/3, 3/0, 0/1, 1/0, 1/2}
			\draw [->] (\from) -- (\to);
			\end{tikzpicture}
		\end{minipage}\hspace*{2cm}
		\begin{minipage}[t]{0.1\textwidth}
			\begin{center}	\begin{tabular}{c c}
					$i$&$\E[X_t^{(i)}]$\\\midrule
					0 & 1\\
					1 & 1\\
					2 & $0.5$\\
					3 & 1
				\end{tabular}
			\end{center}
		\end{minipage}
		\hspace*{1cm}
		\begin{minipage}[t]{0.4\textwidth}	
			\begin{center}$\on{Cov}(X_t^{(i)}, X_t^{(j)})$
				$$
				\begin{blockarray}{ccccc}
				& 0 & 1 & 2 & 3 \\
				\begin{block}{c (c c c c)}
				0 & 1 & $0.5$ & $-0.7$ & $-0.5$\\
				1 & $0.5$ & 1 & 0 & 0\\
				2 & $-0.7$ & 0 & 1 & 0\\
				3 & $-0.5$ & 0 & 0 & 1\\
				\end{block}
				\end{blockarray}
				$$
			\end{center}
		\end{minipage}
	\end{center}
	\begin{center}
		\hspace*{-2.5cm}\begin{minipage}[t]{0.2\textwidth}\vspace*{5mm}
			\begin{tikzpicture}[scale=1, transform shape]
			\tikzstyle{every node} = [circle, fill=gray!30]
			\node (0) at (0, 0) {0};
			\node (1) at +(2,0) {1};
			\foreach \from/\to in {0/1, 1/0}
			\draw [->] (\from) -- (\to);
			\end{tikzpicture}
		\end{minipage} \hspace*{.5cm}
		\begin{minipage}[t]{0.5\textwidth} \vspace*{-5mm}
			\begin{align*}
			\sigma_\gamma^2 &= \frac12 \cdot 2 \var\left( X_t^{(1)} -   X_t^{(0)}\right)\\
			&= 1 + 1 - 2\cdot (0.5) \\
			&= 1.
			\end{align*}
			\vspace*{-5mm}
		\end{minipage}
	\end{center}
	\begin{center}
		\begin{minipage}[t]{0.2\textwidth} 
			\begin{tikzpicture}[scale=1, transform shape]
			\tikzstyle{every node} = [circle, fill=gray!30]
			\node (3) at (0, 0) {3};
			\node (2) at +(2,0) {2};
			\node (0) at +(0,2) {0};
			\foreach \from/\to in {0/2, 2/3, 3/0}
			\draw [->] (\from) -- (\to);
			\end{tikzpicture}
		\end{minipage} \hspace*{1cm}
		\begin{minipage}[t]{0.5\textwidth}\vspace*{-2.8cm}
			\begin{align*}
			\sigma_\gamma^2 &= \frac13\left(\var\left( X_t^{(3)} -   X_t^{(0)}\right)
			+ \var\left( X_t^{(0)} -   X_t^{(3)}\right) \right)\\
			&=\frac23 \left( 1 + 1 - 2\cdot (-0.5) \right)\\
			&= 2.
			\end{align*}
		\end{minipage}
		
		\begin{center}
			\begin{minipage}[t]{0.2\textwidth}
				\begin{tikzpicture}[scale=1, transform shape]
				\tikzstyle{every node} = [circle, fill=gray!30]
				\node (3) at (0, 0) {3};
				\node (2) at +(2,0) {2};
				\node (0) at +(0,2) {0};
				\node (1) at +(2,2) {1};
				\foreach \from/\to in {0/1, 1/2, 2/3, 3/0}
				\draw [->] (\from) -- (\to);
				\end{tikzpicture}
			\end{minipage} \hspace*{1cm}
			\begin{minipage}[t]{0.5\textwidth}\vspace*{-2.8cm}
				\begin{align*}
				\sigma_\gamma^2 &= \frac14\Bigl(\var\left( X_t^{(1)} -   X_t^{(0)}\right)
				+ \var\left( X_t^{(3)} -   X_t^{(1)}\right) \\
				&\hspace*{1cm} + \var\left( X_t^{(0)} -   X_t^{(3)}\right) \Bigr)\\
				&= 1.5.
				\end{align*}
			\end{minipage}
		\end{center}
	\end{center}
	\caption{Example of calculating $\sigma_*$. We find that $\sigma_*^2=2$, determined by the
		cycle $0\to 2 \to 3\to 0$. 
		The covariances are arbitrary --- they would be determined by the random variables associated with these nodes --- and have been chosen to illustrate various possibilities for the impact of covariance on $\sigma_*^2$.}
	\label{F:sigma_star1}
\end{figure}

We also write $\hat\sigma^2:= \max_i \var(X_t^{(0)}-X_t^{(i)})$.

When we convert a diapause model to a migration model, as described in section \ref{sec:diapause}, treating the age classes as sites,
the migration graph $\M$ is simply the cyclic graph $0\to 1 \to \cdots \to d-1 \to 0$.
The variance corresponding to each shift of one step along this graph involves a variance of log ratios of survival or birth rates one age-class apart, summed over one whole generation.
This variance is the same for every shift.
Thus the quantity $\sigma_*^2$ defined at the end of section \ref{sec:migration} is
\begin{equation} \label{E:sigma_diapause}
	\sigma_*^2=\var \left( \log \frac{S_t^{(0)}}{B_t}  \right)+ \var \left( \log \frac{B_t}{S_t^{(d-2)}}  \right)+ \sum_{j=1}^{d-2} \var \left( \log \frac{S_t^{(j)}}{S_t^{(j-1)}}  \right).
\end{equation}

\subsection{Main results} \label{sec:mainresult}
\begin{Thm} \label{T:samerate1}
Suppose there exist sites $i$ and $j$ such that $\mu=\mu_i=\mu_j$ and $X^{(j)}_{t}-X^{(i)}_t$ is
not almost surely zero, that the migration matrices $A_t$ are adapted to the migration graph $\M$, and that $\M$ is connected. 
Then $a$ has modulus of continuity at least $1/\log \epsilon^{-1}$ at $\epsilon=0$. We have
\begin{equation} \label{E:logepsMoC0}
\frac{\sigma_*^2}{2\pi} \le \liminf_{\epsilon\downarrow 0} (\log \epsilon^{-1}) \Bigl( a(\epsilon)-a(0) \Bigr),
\end{equation}
where $\sigma_*^2$ is defined by \eqref{E:definesigma}.

If, in addition, the log growth rates have sub-Gaussian differences  then the modulus of
continuity is $1/\log \epsilon^{-1}$ at $\epsilon=0$. That is,
\begin{equation} \label{E:logepsMoC}
0<\liminf_{\epsilon\downarrow 0} (\log \epsilon^{-1}) \Bigl( a(\epsilon)-a(0) \Bigr)
\leq\limsup_{\epsilon\downarrow 0} (\log \epsilon^{-1}) \Bigl( a(\epsilon)-a(0) \Bigr)
<\infty.
\end{equation}
\end{Thm}

Notice that this is a fairly generic result, as the lower bound does not depend
on any assumptions about the tails. 
The upper bound does depend on the sub-Gaussian
assumption for the logarithms of the matrix entries, meaning that heavy-tailed
distributions --- including, but not exclusively, those that are sub-exponential \cite{jT75},
so {\em a fortiori} entries with polynomial order tail behavior --- could have an even
slower convergence to 0 as $\epsilon$ approaches 0.

\begin{Cor} \label{C:diapause}
In the diapause setting with the birth rate and survival rates $(S_t^{(0)}, S_t^{(1)},\dots,S_t^{(d-2)}, B_t)$ not all deterministic --- that is, at least one entry
has nonzero variance ---
$a$ has modulus of continuity at least $1/\log \epsilon^{-1}$ at $\epsilon=0$. That is,
\begin{equation} \label{E:logepsMoCcor0}
0< \frac{\sigma^2_*}{2\pi d} \le \liminf_{\epsilon\downarrow 0} (\log \epsilon^{-1}) \Bigl( a(\epsilon)-a(0) \Bigr) ,
\end{equation}
where $\sigma_*^2$ is given by the formula \eqref{E:sigma_diapause}.
If $\log B_t$ and $\log S_t^{(j)}$ have sub-Gaussian tails for all $j$ then the modulus of
continuity is $1/\log \epsilon^{-1}$ at $\epsilon=0$. That is,
\begin{equation} \label{E:logepsMoCcor}
0<\liminf_{\epsilon\downarrow 0} (\log \epsilon^{-1}) \Bigl( a(\epsilon)-a(0) \Bigr)
\leq\limsup_{\epsilon\downarrow 0} (\log \epsilon^{-1}) \Bigl( a(\epsilon)-a(0) \Bigr)
<\infty.
\end{equation}
\end{Cor}

\begin{proof}
	The long-term growth rate of the diapause process is $1/d$ times the growth rate $a(\epsilon)$ for the population process resulting from combining $d$ years at a time, as described in section \ref{sec:diapause}.
	If we remove the error term $\epsilon^2 A'_s$ from \eqref{E:diapauseDs} the resulting system satisfies the conditions of Theorem \ref{T:samerate1}, so we need only show that this error term has no effect.
	But since $D_s$ and $A'_s$ are positive, the addition of $\epsilon^2 A'_s$ can only increase $a(\epsilon)-a(0)$, hence the lower bound holds {\em a fortiori}.
	Since the positive parts of the log entries of $A'_s$ are sub-Gaussian, the upper bound on $a(\epsilon)$ still holds, where the migration graph is now the complete graph, and $A_s$ is replaced by $A_s+\delta A'_s$ for an arbitrarily small positive $\delta$.
	It doesn't matter what $\delta$ we choose, because the result that the limsup in \eqref{E:logepsMoC} is finite will hold equally well in any case.
\end{proof}

\subsection{Migration penalties}
Thinking about the migration (or diapause) process that we are purporting to model, it may seem strange that the extra population of size $\epsilon A_t(j,i)N_t(i)$ that appears at site $j$ is not compensated by a corresponding loss at site $i$ --- indeed, perhaps a greater loss, if we suppose that migration is inherently risky or costly.
This is being done purely for simplicity of notation.
We could add random penalties  
proportional to $\epsilon$.
That is, the diagonal terms of $D_t$ would change from $\e^{X_t^{(j)}}$ to $\e^{X_t^{(j)} - \epsilon \Delta_t^{(j)}}$.

The contribution of $\Delta$ to $a$ will
be linear in $\epsilon$, hence negligible in comparison
to the scale $1/\log\epsilon^{-1}$ that we will be
considering. 
For clarity of exposition we will henceforth drop $\Delta$ from
our notation and our proofs, understanding that the
results hold equally well for any $\Delta$ with finite $L^1$ norm.
As these are being conceptualized as ``penalties'', it is natural to suppose that these $\Delta$ terms are positive, but we need not assume this.

\section{Trajectories} \label{sec:Trajectories}
In analyzing the generic migration problem in \cite{migration2018}, a central role was played by the
enumeration of ``excursions'' away from the optimal-growth site. The vast majority of the population will have an ancestry that spent nearly all of its time at that site, but made rare excursions to other sites at times when those
happened to have periods of exceptionally large growth.

In the current setting the optimal ancestries will have divided their time more or less equally among the optimal sites. There is no home base from which to count excursions. What we need to enumerate are ``trajectories'', which will simply be
paths in the migration graph $\M$. The set of all trajectories
of length $T$ will be denoted $\F_{T}$, and the set
of trajectories that start at site $i$ and end at site $j$ will be $\F_T(i,j)$. The set of changepoints of a trajectory $f$ will be denoted
$$
K(f):=\{t\, :\, f_{t}\ne f_{t+1}\}.
$$
We write $\F_{T,k}$ for the set of trajectories with exactly $k$ changepoints,
and we have ${\binom{T}{k}\le\#\F_{T,k}\le d^{k+1} \binom{T}{k}}$.
We endow $\F_{T}$ with the $L^{2}$ norm $\|\cdot\|_{2}$, defined to
be the square root of the Hamming distance (the number of timepoints at which the trajectories
are not equal). The {\em null trajectory}
$f^{(0)}$ will denote the path that stays at 0 for all $T$ steps.

\nc{\ff}{\mathbf{f}}
We then have the random variables
$$
Z_{f}:=f[\bX,\bA]:=\sum_{t\in K(f)} \log A_{t}(f_{t+1},f_{t})
   +\sum_{t\in \{0,\dots,T-1\}} \left( X_{t}^{(f_{t})} - X_t^{(0)} \right).
$$
Here $0$ is assumed to be a site with maximum
mean log growth rate, but is otherwise arbitrary.
(We will use the $Z_{f}$ notation for brevity when there is no need
to emphasize the dependence on $\bX$ and $\bA$.)

\begin{lemma} \label{L:allterms2}
\begin{equation} \label{E:allterms2}
\log R_{T}(\epsilon)_{00}=\sum_{t=0}^{T-1}X_{t}^{(0)}
  +\log \Bigl(1+  \sum_{f\in \F_{T}(0,0)\setminus\{f^{(0)}\}} \e^{ f[\bX,\bA] }\epsilon^{\# K(f)}\Bigr),
\end{equation}
where $f^{(0)}$ is the null trajectory. Thus
\begin{equation} \label{E:allterms3}
\begin{split}
\liminf_{T\to\infty} & \: T^{-1} \max_{f\in \F_{T}} \Bigl\{ f[\bX,\bA] - \# K(f)\log\epsilon^{-1} \Bigr\} 
\le a(\epsilon)-a(0)\\
&\le  \limsup_{T\to\infty} T^{-1} \Bigl( \log  \#\F_{T} +\max_{f\in\F_{T}} \Bigl\{f[\bX,\bA ]  - \# K(f)\log\epsilon^{-1}\Bigr\}\Bigr)
\end{split}
\end{equation}
\end{lemma}

\begin{proof}
	We have, by definition,
	\begin{equation} \label{E:prodhe}
	R_{T}(\epsilon)_{00}=\sum_{(f_{0},\dots,f_{T})} \prod_{t=0}^{T-1} D_{t}(f_{t+1},f_{t}),
	\end{equation}
	where the summation is over $(f_{0},\dots,f_{T})\in \{0,\dots,d-1\}^{T+1}$ with
	$f_{0}=f_T = 0$. Note that we may restrict the summation to $(T+1)$-tuples such
	that $D_{t}(f_{t+1},f_{t})>0$, which will only be true when $(f_{t},f_{t+1})$ is an edge of $\M$.
	These are the trajectories in $\F_T(0,0)$.
	
	We have $D_{t}(0,0)=\e^{X_{t}^{(0)}}$. Thus, we may write the log of the expression in \eqref{E:prodhe} as
	\begin{align*} 
		&\log R_{T}(\epsilon)_{00} = \sum_{t=0}^{T-1} X_t^{(0)} +
		\log\left( \sum_{f\in \F_T(0,0)} \prod_{t=0}^{T-1} \frac{D_t(f_{t+1},f_t)}{D_t(0,0)} \right)\\
		&= \sum_{t=0}^{T-1} X_t^{(0)} +
		\log\left( 1 + \sum_{f\in \F_T(0,0)\setminus\{f^{(0)}\}} \exp\left\{
		\sum_{t=0}^{T-1} \log D_t(f_{t+1},f_t) - X_t^{(0)} \right\} \right) .
	\end{align*}
	The definition of $D_t$ immediately yields the expression $f[\bX,\bA] - \# K(f) \log \epsilon^{-1}$,	completing the proof.
\end{proof}

\section{Proof of the lower bound}  \label{sec:sameLB}
Intuitively, the argument is simply that we can find a subset of cycles, of determined length, such that the maximum cumulative increase (relative to the null trajectory) has positive expectation, bounded below by an appropriate function of $\epsilon$ and of the length.
By repeating these cycles indefinitely we obtain a sequence of i.i.d.\ variables whose sum bounds the maximum (over all trajectories) from below.
The almost sure lower bound is then a simple consequence of the law of large numbers.

\subsection{The cycle and its admissible sites}
Let $\gamma=(0,\gamma_1,\dots, \gamma_{k-1}, 0)$
be a cycle in $\M$ that maximizes $\sigma_*^2$.
For convenience we will extend the definition of $\gamma$ to $\gamma_t$ for $t\in\R_+$ by $\gamma_t:=\gamma_{\lfloor t \rfloor\, \operatorname{mod}\, k}$.

Following the definition of $\tilde\gamma$ in section \ref{sec:migration}, we delete from $\gamma$ every site that does not have the maximum mean log growth rate; and we then delete, successively, any remaining site whose log growth rate differs almost surely by a constant from that of its predecessor among the sites still remaining.
The sites that survive we call {\em admissible}, and we write
$$
\gamma'=\bigl(\gamma_{j_0},\gamma_{j_1},\dots,\gamma_{j_{k'-1}}\bigr), \qquad 0=j_0<j_1<\cdots<j_{k'-1}<k,
$$
for the subpath they form, setting $j_{k'}:=k$, so that $\gamma_{j_{k'}}=\gamma_0=0$.
(The site $0$ is never deleted: it is optimal by assumption, and the second rule deletes only successors.)
We write
$$
r_i:= j_{i+1}-j_i-1 \ge 0
$$
for the number of inadmissible sites lying between successive admissible sites, so that
$k=k'+\sum_{i=0}^{k'-1} r_i$, and $\gamma$ may be written as
\begin{multline*}
	0=\gamma_{j_0},\gamma_{j_0+1},\dots,\gamma_{j_0+r_0},\\
	\gamma_{j_1},\gamma_{j_1+1},\dots,\gamma_{j_1+r_1},\,
	\dots,\\
	\gamma_{j_{k'-1}},\gamma_{j_{k'-1}+1},\dots,\gamma_{j_{k'-1}+r_{k'-1}},\,
	\gamma_{j_{k'}}=0.
\end{multline*}
We write
$$
\sigma_i^2:= \var\bigl(X_t^{(\gamma_{j_{i+1}})} - X_t^{(\gamma_{j_{i}})} \bigr)
$$
for the variance associated with the passage from one admissible site to the next.
The second deletion rule removes only sites whose log growth rate is almost surely equal to that of the preceding surviving site, so it changes none of these variances; consequently
\begin{equation} \label{E:sumsigma}
	\sum_{i=0}^{k'-1} \sigma_i^2 = k \sigma_*^2 ,
\end{equation}
with the number of sites of the {\em whole} cycle, $k$, appearing on the right, exactly as in the definition \eqref{E:definesigma}.
By construction $\sigma_i>0$ for every $i$, and we write $\sigma_{\min}:=\min_i \sigma_i>0$.

The reader may find it helpful throughout to keep in mind the special case in which every site of $\gamma$ is admissible, so that $k'=k$ and every $r_i=0$; this is the situation pictured in Figure \ref{F:base_trajectory}.

\begin{figure}[ht]
	\hspace*{-1cm}\begin{tikzpicture}[scale=.4]
		\foreach \y in {0,...,3}{
			\draw[very thick] (0,-2.5*\y) -- (33,-2.5*\y);
			\draw (-1,-2.5*\y) node{\y};
			\foreach \x in {0,...,33}
			\draw (\x , 0-2.5*\y) -- (\x , 0.4-2.5*\y);
		}
		\foreach \x in {0,...,6}
		\draw (5*\x, -8.5) node {%
			\pgfmathparse{5*\x}
			\pgfmathprintnumber{\pgfmathresult}};
		\draw[red,line width=2pt] (0,0.2) -- (9,0.2) -- (10,-4.8) -- (17,-4.8) -- (18,-7.3) -- (23,-7.3) -- (24, 0.2) -- (33, 0.2);
		\draw[green,line width=2pt] (0,-0.2) -- (6,-0.2) -- (7,-5.2) -- (11,-5.2) -- (12,-7.7) -- (22,-7.7) -- (23, -0.2) -- (27, -0.2) -- (28, -5.2) -- (33, -5.2);
	\end{tikzpicture}
	\caption{The reference sequence $\beta$ (red) and one element
		of $\F_*$ (green) on the cycle $\gamma=(0,2,3,0)$,
		based on the cycle defined in the example
		of Figure \ref{F:sigma_star1}, with ${\ell_0=10}$, ${\ell_1=8}$, ${\ell_2=6}$,
		so that $L_1=10$, $L_2=18$ and $L=24$.
		We are assuming for purposes of this example that 0, 2, and 3 all
		have the maximum mean log growth rate, and that the associated variances
		$\sigma_i^2$ are all positive; every site of $\gamma$ is then admissible, so that
		$\gamma_{j_0}=0$, $\gamma_{j_1}=2$, $\gamma_{j_2}=3$, with $k'=k=3$ and every $r_i=0$.
		In this special case $\beta$ is itself a path in $\M$, though never an element of
		$\F_*$, since it leaves each site only at the end of the corresponding block.
		If the green trajectory is supposed to yield the maxima
		defining $Z(i)$ in \eqref{E:Zi} then, omitting the passage penalties $\Lambda_i(s)$,
		$$
		Z(0)=\sum_{t=7}^9 \left(X_t^{(2)} - X_t^{(0)}\right),\qquad
		Z(1)=\sum_{t=12}^{17} \left(X_t^{(3)} - X_t^{(2)}\right),
		$$
		$$
		Z(2)= \left(X_{23}^{(0)} - X_{23}^{(3)}\right), \qquad
		Z(3)=\sum_{t=28}^{33} \left(X_t^{(2)} - X_t^{(0)}\right).
		$$
		The figure shows one complete journey around $\gamma$, of length $L=24$, followed
		by the first block of the next; $Z(3)$ has the same distribution as $Z(0)$.
	}
	\label{F:base_trajectory}
\end{figure}

\begin{figure}[ht]
	\hspace*{-1cm}\begin{tikzpicture}[scale=.4]
		\foreach \y in {0,...,3}{
			\draw[very thick] (0,-2.5*\y) -- (33,-2.5*\y);
			\draw (-1,-2.5*\y) node{\y};
			\foreach \x in {0,...,33}
			\draw (\x , 0-2.5*\y) -- (\x , 0.4-2.5*\y);
		}
		\foreach \x in {0,...,6}
		\draw (5*\x, -8.5) node {%
			\pgfmathparse{5*\x}
			\pgfmathprintnumber{\pgfmathresult}};
		\draw[red,dashed,line width=2pt] (0,0.2) -- (9,0.2) -- (10,-7.3) -- (16,-7.3) -- (17, 0.2) --
		(26, 0.2) -- (27, -7.3) -- (33, -7.3);
		\draw[green,line width=2pt] (0,-0.2) -- (5,-0.2) -- (6,-5.2) -- (7,-7.7) -- (15,-7.7) -- (16, -0.2) --
		(18, -0.2) -- (19, -5.2) -- (20, -7.7) -- (32, -7.7) -- (33, -0.2);
	\end{tikzpicture}
	\caption{The reference sequence $\beta$ (red, dashed) and one element
		of $\F_*$ (green) on the same cycle $\gamma=(0,2,3,0)$ as in
		Figure \ref{F:base_trajectory}, but now with site 2 having less than the
		maximum mean log growth rate, as in the example of Figure \ref{F:sigma_star1}.
		Site 2 is therefore inadmissible, and the admissible subpath is
		$\gamma'=(\gamma_{j_0},\gamma_{j_1})=(0,3)$, with $k'=2$, $r_0=1$ and $r_1=0$,
		so that $k=k'+\sum_i r_i=3$. Taking $\ell_0=10$ and $\ell_1=6$ gives
		$L_1=\ell_0+r_0=11$ and $L=17$.
		Trajectories in $\F_*$ pass through site 2 in a single time unit, at a moment
		of their own choosing, whereas $\beta$ does not visit site 2 at all: it occupies
		$\gamma_{j_0}=0$ for the first $\ell_0$ steps of block 0 and $\gamma_{j_1}=3$ for the
		remaining $r_0=1$, so that every $\beta_t$ is an admissible site with mean log
		growth rate $\mu$. It is drawn dashed as a reminder that it is a bookkeeping
		sequence and not a path in $\M$.
		If the green trajectory is supposed to yield the maxima defining $Z(i)$
		in \eqref{E:Zi}, then it leaves site 0 at $s=6$, so that
		$$
		\Lambda_0(6)= \log A_{6}(2,0) + \log A_{7}(3,2) + \bigl( X_{6}^{(2)} - X_{6}^{(3)}\bigr),
		$$
		exhibiting the two migration penalties of the passage together with the growth
		rate incurred while traversing site 2, measured against the destination, and
		$$
		Z(0)=\Lambda_0(6)+\sum_{t=6}^9 \left(X_t^{(3)} - X_t^{(0)}\right),\qquad
		Z(1)=\Lambda_1(5)+ \left(X_{16}^{(0)} - X_{16}^{(3)}\right),
		$$
		$$
		Z(2)=\Lambda_2(2)+ \sum_{t=19}^{26} \left(X_t^{(3)} - X_t^{(0)}\right), \qquad
		Z(3)=\Lambda_3(5)+ \left(X_{33}^{(0)} - X_{33}^{(3)}\right).
		$$
		The figure shows exactly two journeys around $\gamma$, since $2L=34$.
	}  \label{F:base_trajectory2}
\end{figure}

\subsection{The reference sequence and the class $\F_*$}
Fix a cyclically repeating sequence of positive integers $\ell_0,\dots,\ell_{k'-1},\ell_{k'},\dots$, where $\ell_i=\ell_{i'}$ for $i\equiv i' \operatorname{mod} k'$, and define
$$
L_i:= \sum_{i'< i} \bigl(\ell_{i'}+r_{i'}\bigr).
$$
For any non-negative integer $t$ we define $\langle t\rangle$ to be the unique $i$ such that $L_i\le t < L_{i+1}$;
we refer to the times $L_i\le t<L_{i+1}$ as {\em block} $i$.
We also write $L:= L_{k'}$, the length of one journey around the cycle, and $\ell_*:= \min\ell_i$.
Note that $L=\sum_{i<k'}\ell_i + (k-k')$.

The maximum over $f\in \F_T$ described in \eqref{E:allterms3}
will be bounded below by the maximum over a more manageable set of trajectories $\F_*\subset \F_T$.
These are the trajectories that are at $\gamma_{j_i}$ at each of the times $L_i$, leave
$\gamma_{j_i}$ at some time $L_i+s$ with $0\le s\le \ell_i-r_i-1$, traverse the $r_i$
intervening inadmissible sites as rapidly as possible --- one time unit in each ---
and then remain at $\gamma_{j_{i+1}}$ until time $L_{i+1}$.
Thus, for example, if $\gamma_{j_0}=0$ and $\gamma_{j_1}=1$ are adjacent in $\M$ (so that $r_0=0$)
a trajectory in $\F_*$ may move from 0 to 1 at any time
between $t=0$ and $t=\ell_0-1$; once arrived, it remains
at least until time $\ell_0-1$.
To these we adjoin the trajectories that remain at 0 from time $jL$ to $(j+1)L$, and so skip one journey around the cycle $\gamma$.
(This is made necessary by the choice to allow entries in $A$ to be arbitrarily small --- and indeed, possibly zero with positive probability.
This means that we can't guarantee that there will be any favorable time to make one step in the cycle, without annihilating the population, though the probability that some appropriate times will be available converges to 1 as $L$ goes to $\infty$.)
We write $\F_{*,T}$ for
the trajectories in $\F_*$ of length $T$. An example is pictured in Figure \ref{F:base_trajectory}, and one with $r_0=1$ in Figure \ref{F:base_trajectory2}.

Every trajectory in $\F_{*,mL}$ makes exactly $mk$ changes of site, since each of the $mk'$ passages
comprises $r_i+1$ steps, and $\sum_{i=0}^{k'-1}(r_i+1)=k$.

\subsection{Passage penalties}
It will be convenient to compare trajectories against the sequence of sites
\begin{equation} \label{E:betadef}
	\beta_t := \begin{cases}
		\gamma_{j_{\langle t\rangle}} & \text{ if } L_{\langle t\rangle}\le t< L_{\langle t\rangle}+\ell_{\langle t\rangle},\\
		\gamma_{j_{\langle t\rangle+1}} & \text{ if } L_{\langle t\rangle}+\ell_{\langle t\rangle}\le t< L_{\langle t\rangle+1},
	\end{cases}
\end{equation}
which spends the whole of block $i$ at the two admissible sites $\gamma_{j_i}$ and $\gamma_{j_{i+1}}$, ignoring the intervening passage.
Note that $(\beta_t)$ is in general not a path in $\M$; it is used only as a bookkeeping device, and its
merit is that $\E\bigl[X^{(\beta_t)}_t\bigr]=\mu$ for every $t$.

For integers $i$ we define
\begin{equation} \label{E:Zi}
	Z(i):= \max_{0\le s \le \ell_i -r_i-1 }\Bigl\{ \Lambda_i(s) +
	\sum_{v=0}^{\ell_{i}-s-1} \Bigl(
	X^{(\gamma_{j_{i+1}})}_{L_i+s+v} - X^{(\gamma_{j_{i}})}_{L_i+s+v} \Bigr) \Bigr\},
\end{equation}
where the sum accounts for the gain over block $i$ of a trajectory leaving
$\gamma_{j_i}$ at time $L_i+s$, computed as though its arrival at $\gamma_{j_{i+1}}$
were instantaneous, and
\begin{equation} \label{E:Lambda}
	\begin{split}
		\Lambda_i(s):= \sum_{v=0}^{r_i} \log A_{L_i+s+v}\bigl(\gamma_{j_i+v+1},\gamma_{j_i+v}\bigr)
		+ \sum_{v=0}^{r_i-1} \Bigl( X^{(\gamma_{j_i+v+1})}_{L_i+s+v} - X^{(\gamma_{j_{i+1}})}_{L_i+s+v} \Bigr)
	\end{split}
\end{equation}
is the correction for the time actually spent in transit: the migration penalties
along the passage, together with the growth foregone, relative to the destination,
while traversing the inadmissible sites.
Both sums in $\Lambda_i(s)$ have non-positive mean, the second
because $\gamma_{j_{i+1}}$ is admissible and so attains the maximum mean log growth
rate $\mu$. (When $r_i=0$ the second sum is empty and $\Lambda_i(s)$ reduces to the
single term $\log A_{L_i+s}(\gamma_{j_{i+1}},\gamma_{j_i})$.)

Since $\gamma$ is a path in $\M$ we have by our assumptions that
$A_{L_i+s+v}(\gamma_{j_i+v+1},\gamma_{j_i+v})$ is not almost surely 0 for any $v$, and since the $X_t$ have finite
expectation the variables in \eqref{E:Lambda} are almost surely finite; hence we may find real numbers
$A_*$ and $p\in (0,1/8]$ such that $\P\{\Lambda_{i}(s) \ge -A_*\}\ge \sqrt{8p}$
for all $i$ and $s$.
The variables $\Lambda_i(s)$ and $\Lambda_i(s')$ are generated from disjoint collections of random variables $X_t$ and $A_t$ when $|s-s'|>r_i$.
We assume $\ell_*\ge 2k$ (which will hold for the choice of $\ell_i$ we fix below in \eqref{E:elli} when $\epsilon$ is small enough) and write $n_i=\lfloor \ell_i/k\rfloor$.
We then restrict attention to the $n_i$ departure times $s=0,k,2k,\dots,k(n_i-1)$.
The separations are all $k$, which is greater than $r_i$ --- the number of inadmissible sites included in the path must be smaller than the length of the whole path --- so the corresponding events $\{\Lambda_i(s)\ge -A_*\}$ are independent.
Define
$$
\eu{A}_i := \Bigl\{ 1-\tfrac{s}{\ell_i} \; : \; s=k\iota \text{ for some } 0\le \iota< n_i, \text{ and } \Lambda_i(s)\ge -A_* \Bigr\}.
$$
These may be understood to be a random set of times when the passage from $\gamma_{j_i}$ to $\gamma_{j_{i+1}}$ may be begun without excessive penalty.
For integers $i\ge 0$ we define $\eu{B}_i$ to be the event that $\#\eu{A}_j\ge \frac{p}{2}n_j $ for all $j$ with $\lfloor j/k' \rfloor = \lfloor i/k' \rfloor$, and we write $n_*:=\min_i n_i$.
Note that this definition implies that $\mathcal{B}_{jk'}, \mathcal{B}_{jk'+1},\dots, \eu{B}_{jk'+k'-1}$ all denote precisely the same event.
For $i$ in distinct blocks of $k'$ the events $\eu{B}_{i}$ are independent, and by Hoeffding's inequality
\begin{equation}
	\label{E:Bprob}
	\P (\eu{B}_{i})  \ge 1- k' \e^{-p n_*} .
\end{equation}

\subsection{Reduction to a sum of i.i.d.\ variables}
For any fixed $T$ and $(\ell_i)$, the collection of random variables ${\{Z(i): 0\le i \le \langle T \rangle \}}$
is independent, and for each positive integer $m$
\begin{align}
	\max\bigl\{ Z_{f}\, :\, f\in \F_{m L, mk} \bigr\}
	&\ge \max\bigl\{ Z_{f}\, :\, f\in \F_{*,m L} \bigr\} \notag\\
	& \ge  \sum_{i=0}^{k' m-1} Z( i) \indic_{\eu{B}_i} \, - \,
	\sum_{t=0}^{mL-1} \left( X^{(0)}_{t} - X^{(\beta_{t})}_{t} \right) \indic_{\eu{B}_{\langle t \rangle}}.\label{E:maxZf}
\end{align}
(Here we are making use of the option to ``sit out'' a cycle when there are too few good options for making one of the passages --- defined by the occurrence of the event $\eu{B}_i$ --- and instead squat the whole period on the null trajectory, which we have included in the definition of $\F_{*,m L}$.)
In addition, $Z(i)$ and $Z(i')$ have the same distribution when $i \equiv i' \on{mod}\, k'$.
For $0\le i\le k'-1$ we define
\begin{equation} \label{E:Gammal}
	\Gamma_{\ell,i}:=\ell_i^{-1/2}\E\bigl[ Z(i) \indic_{\eu{B}_i}\bigr].
\end{equation}
Since $\beta_t$ takes values only among the admissible sites, all of which have mean log growth rate $\mu$,
Lemma \ref{L:Yml} applies with no drift term, and it follows from \eqref{E:maxZf} and then the Strong Law of Large Numbers that for fixed $(\ell_i)$,
\begin{equation} \label{E:maxZf2}
	\begin{split}
		\liminf_{m\to\infty}&m^{-1}\max\bigl\{ Z_{f}\, :\, f\in \F_{mL , mk} \bigr\} \\
		& \ge \sum_{i=0}^{k'-1} \ell^{1/2}_i \Gamma_{\ell,i}  - \hat\sigma \sqrt{k'L}\e^{-p n_*/2}
		\text{ a.s.}
	\end{split}
\end{equation}

We now fix
\begin{equation} \label{E:elli}
	\ell_i = \left\lceil 2\pi \left( \frac{\sigma_i}{ \sigma_*^2} \log \epsilon^{-1} \right)^2 \right\rceil,
\end{equation}
so that, by \eqref{E:sumsigma} and since $L<\sum_{i<k'}\ell_i+k$,
\begin{align}
	\label{E:kLbound}	&\frac{\sigma_*^2}{2\pi \log^2 \epsilon^{-1}} \ge
	\frac{k}{L} \ge
	\frac{\sigma_*^2}{2\pi \log^2 \epsilon^{-1}} - \left( \frac{\sigma_*^2}{2\pi \log^2 \epsilon^{-1}}  \right)^2
	\, ,\text{ and}\\
	\label{E:nstarbound}	&n_*\ge \frac{\pi}{2k} \Bigl(\frac{\sigma_{\min}}{\sigma_*^2} \Bigr)^2  \log^2 \epsilon^{-1} ,
\end{align}
so that $\e^{-p n_*/2}$ decays faster than any power of $\epsilon$, and $k/L=O(\log^{-2}\epsilon^{-1})$.

For any $T\ge mL$ we then have
\begin{align*}
	T^{-1}\max\bigl\{ & Z_{f}-\# K(f) \log\epsilon^{-1}\, :\, f\in \F_{T} \bigr\}\\
	& \ge T^{-1}\max\bigl\{ Z_{f}-\# K(f)\log\epsilon^{-1}\, :\, f\in \F_{mL} \bigr\}\\
	&\ge T^{-1}\max\bigl\{ Z_{f}\, :\, f\in \F_{mL, mk} \bigr\} -T^{-1} mk \log\epsilon^{-1}, \\
	&\hspace*{3cm} \text{ since }
	\# K(f)=k m \text{ for any }f\in \F_{mL,mk}\\
	& \ge  \left(1-\frac1m\right) L^{-1} m^{-1} \max\bigl\{ Z_{f}\, :\, f\in \F_{*, mL} \bigr\} -\frac{k}{L} \log\epsilon^{-1}
\end{align*}
where $m=\lfloor T/L\rfloor$.

\subsection{The Brownian-motion limit}
Define for $u\in [0,1]$
$$
W_{u}^{(i)}:= \frac{1}{\sqrt{\ell_i}\sigma_i}  \sum_{(1-u)\ell_i \le r < \ell_i } \left(
X^{(\gamma_{j_{i+1}})}_{L_i + r} - X^{(\gamma_{j_{i}})}_{L_i + r} \right).
$$
Then, writing $u=1-s/\ell_i$,
\begin{align*}
	Z(i) &= \sqrt{\ell_i}\sigma_i \max_{0\le s \le \ell_i-r_i-1} \Bigl( W_{1-s/\ell_i}^{(i)} +  \ell_i^{-1/2} \sigma_i^{-1} \Lambda_i(s) \Bigr)\\
	& \ge \sqrt{\ell_i}\sigma_i \Bigl( \max_{u \in \eu{A}_{i}} W_u^{(i)} \Bigr) -  \sqrt{\ell_i}\sigma_i \Bigl( \max_{u \in [0,1]} W_u^{(i)}  \Bigr) \indic_{\eu{B}^\complement_i} -  A_* ,
\end{align*}
where $\eu{A}_{i}$ is nonempty on the event $\eu{B}_i$.

$\eu{A}_i$ is a Bernoulli process on a set of $n_i$ equally spaced points of $(0,1]$, retaining each with probability $\ge \sqrt{8p}$.
The longest run of consecutive omitted points among $n_i$ independent trials is $O_{\mathrm{P}}(\log n_i)$, so the maximum gap in $\eu{A}_i$ is $O_{\mathrm{P}}(k\log n_i/\ell_i)$, which converges to 0 in probability as $\epsilon\downarrow 0$.
By Lemma \ref{L:invariance}, since $\ell_i\to\infty$ when $\epsilon\downarrow 0$,
$$
	\lim_{\epsilon\downarrow 0} \E\Bigl[ \max_{u \in \eu{A}_{i}} W_u^{(i)} \Bigr]=\sqrt{2/\pi};
$$
and by Lemma \ref{L:invariance} and the Cauchy--Schwarz inequality
$$
  \E\Bigl[ \max_{u \in [0,1]} W_u^{(i)}  \indic_{\eu{B}^\complement_i}\Bigr] \le  \E\Bigl[ \max_{u \in [0,1]} (W_u^{(i)} )^2\Bigr]^{1/2} \E\Bigl[\indic_{\eu{B}^\complement_i}\Bigr]^{1/2} \le 2\sqrt{k'}\e^{-pn_*/2}.
$$
It follows that for any $\delta>0$ we may find $\epsilon_0$ such that for $0<\epsilon <\epsilon_0$ (hence $n_*$ sufficiently large)
\begin{equation} \label{E:donskerlimit}
	\Gamma_{\ell,i}\ge \sqrt{2/\pi} \cdot \sigma_i - 2\sqrt{k'}\sigma_i \e^{-p n_*/2} -\delta\cdot \frac{\sum_{j=0}^{k'-1} \sigma_j}{k \sqrt{2\pi}}
	-\ell_i^{-1/2} A_* \, .
\end{equation}
Recalling \eqref{E:sumsigma}, and hence by the Cauchy--Schwarz
inequality also that \\
${\sum_{i=0}^{k'-1} \sigma_i \le \sqrt{k'k}\,\sigma_*\le k \sigma_*}$, we then have
\begin{align*}
	\sum_{i=0}^{k'-1} \sqrt{\ell_i} \Gamma_{\ell,i} &\ge \frac{\sqrt{2}}{\sqrt{\pi}}\cdot \sqrt{2\pi}\log \epsilon^{-1}
	\sum_{i=0}^{k'-1} \sigma_i \cdot \frac{\sigma_i}{\sigma_*^2}\\
	&\qquad \qquad - \delta \cdot \frac{\sum_{j=0}^{k'-1} \sigma_j}{k \sqrt{2\pi}} \left(1+ \sqrt{2\pi}\log\epsilon^{-1} \frac{\sum_{i=0}^{k'-1} \sigma_i}{\sigma_*^2}\right) - k' A_*\\
	&\ge k\log\epsilon^{-1} \left( 2 - \delta\right) - \frac{\sigma_*}{\sqrt{2\pi}} \delta - kA_* \, ,
\end{align*}
By \eqref{E:maxZf2} it follows that
\begin{equation} \label{E:aLbound}
	\begin{split}
		a(\epsilon)&-a(0)
		\ge \lim_{T\to\infty}T^{-1}\max\bigl\{ Z_{f}- \# K(f)\log\epsilon^{-1}\, :\, f\in \F_{T} \bigr\}\\
		& \ge
		L^{-1} \Bigl(\sum_{i=0}^{k'-1} \sqrt{\ell_i} \Gamma_{\ell,i} - \hat\sigma \sqrt{k'L}\e^{-p n_*/2} - k\log \epsilon^{-1}  \Bigr) \\
		&\ge \frac{k}{L}(1-\delta) \log\epsilon^{-1} - \hat\sigma \sqrt{\frac{k}{L}}\e^{-p n_*/2} -  \frac{\sigma_* \delta}{\sqrt{2\pi} L } -\frac{ kA_*}{L}\\
		&\ge \frac{\sigma_*^2}{2\pi \log \epsilon^{-1}}
		\left( 1- \frac{\sigma_*^2}{2\pi \log^2 \epsilon^{-1}}  \right)(1-\delta) -
		\hat\sigma \e^{-\frac{\pi p}{4k} (\sigma_{\min}/\sigma_*^2)^2 \log^2 \epsilon^{-1}}\\
		&\hspace*{3cm} -  \frac{\sigma_*^3 \delta + \sqrt{2\pi} k \sigma_*^2 A_*}{(2\pi)^{3/2} k } \cdot \frac{1}{\log^2 \epsilon^{-1}}.
	\end{split}
\end{equation}
Since $\delta>0$ is arbitrary,
$$
\liminf_{\epsilon\downarrow 0} \bigl( a(\epsilon)-a(0)\bigr) \log\epsilon^{-1}
\ge \frac{\sigma_*^2}{2\pi} .
$$

\begin{lemma} 	\label{L:Yml}
	With definitions as above, and recalling ${\hat\sigma^2 :=\max_i \var\bigl(X_t^{(0)}-X_t^{(i)}\bigr)}$,
	$$
	\Bigl|\E\Bigl[\sum_{t=0}^{mL-1} \left( X^{(0)}_{t} - X^{(\beta_{t})}_{t} \right) \indic_{\eu{B}_{\langle t \rangle}} \Bigr] \Bigr|
	\le m\sqrt{k'L}\, \hat\sigma\, \e^{-p n_*/2} .
	$$
\end{lemma}
\begin{proof}
	The random variables in the sum repeat in distribution at intervals of $L$, so it
	suffices to show the result for $m=1$; and since $\langle t\rangle\le k'-1$ for $t<L$,
	we have $\eu{B}_{\langle t\rangle}=\eu{B}_0$ throughout, so that the indicator may be taken
	outside the sum.
	
	Write $S:=\sum_{t=0}^{L-1}\bigl( X^{(0)}_{t} - X^{(\beta_{t})}_{t}\bigr)$.
	Every $\beta_t$ is an admissible site, hence has mean log growth rate $\mu$, so $\E[S]=0$ and
	therefore $\E\bigl[S\indic_{\eu{B}_{0}}\bigr]=-\E\bigl[S\indic_{\eu{B}_{0}^\complement}\bigr]$.
	Applying the Cauchy--Schwarz inequality and \eqref{E:Bprob}, we have
	\begin{align*}
		\Bigl|	\E\bigl[S \indic_{\eu{B}_{0} } \bigr] \Bigr|
		&\le 	\E\bigl[S^2\bigr]^{1/2} \P\bigl(\eu{B}_{0}^\complement \bigr)^{1/2} \\
		&\le \Bigl(\sum_{t=0}^{L-1} \var\bigl(X_t^{(0)}-X_t^{(\beta_t)}\bigr) \Bigr)^{1/2} \cdot \sqrt{k'}\e^{-p n_*/2}\\
		&\le  \sqrt{k'L}\, \hat\sigma\, \e^{-p n_*/2}.
	\end{align*}
\end{proof}

\begin{lemma} \label{L:invariance}
	Let $\xi_0^{(n)},\dots,\xi_{n-1}^{(n)}$ be a triangular array of independent random variables with
	mean 0 and variance 1. Let $\eu{A}_n$ be point processes on $[0,1]$ such that
	the size of the maximum gap $G_n:= \sup_{0<s<t<1} \{t-s \, : \, \eu{A}_n \cap [s,t] = \emptyset \}$ converges to 0 in probability.
	Define for $u \in [0,1]$
	$$
		W_u := 	n^{-1/2} \sum_{0\le i < un } \xi_i^{(n)}.
	$$
	Then $\sup_{u\in \eu{A}_n} W_u$ converges in distribution to the maximum of a standard
	Brownian motion over $[0,1]$, and
	\begin{equation} \label{E:Wmax}
		\lim_{n\to\infty} \E\left[\sup_{u\in \eu{A}_n} W_u \right] =\sqrt{2/\pi} .
	\end{equation}
	Also,
	\begin{equation} \label{E:Wmax2}
		\E[\sup_{u\in [0,1]} W_u^2 ]\le 4.
	\end{equation}
\end{lemma}

\begin{proof}
	By Donsker's invariance principle ({\em cf.} Theorem 2.4.4 of \cite{EKM97})
	$(W^{(n)}_{u})_{0\le u\le 1}$ converges weakly (in supremum) to a Brownian motion 
	$(\omega_{u})_{0\le u\le 1}$. Thus $W_*^{(n)}:= \sup_{u\in [0,1]} W_u^{(n)}$ converges
	in distribution to $\sup_{u\in [0,1]} \omega_u$. We now show that 
	$W_*^{(n)} - \sup_{u\in \eu{A}_n} W_u^{(n)}$ converges in probability to 0.
	We have
	$$
		0\le W_*^{(n)} - \sup_{u\in \eu{A}_n} W_u^{(n)} \le \sup_{u,v\in [0,1)\, : \, |u-v|\le G_n} \left| W_v^{(n)} - W_u^{(n)} \right|
	$$
	
For any $\delta>0$ let
$$
	S_\delta:= \frac{\sup_{u,v\in [0,1]\, : \, |u-v|\le \delta} |\omega_v - \omega_u|}{\sqrt{\delta\log \delta^{-1}}}.
$$
By L\'evy's Theorem on the modulus of continuity of Brownian motion \cite[Theorem 9.25]{KS14}
$\limsup_{\delta \downarrow 0} S_\delta =\sqrt{2}$ almost surely. 
It follows that ${\lim_{\delta\downarrow 0} \P\{S_\delta \ge 2 \} =0}$. 
Thus we may find, for any positive integer $m$, a $\delta_m>0$ such that
\begin{align*}
	\P \Bigl\{  \sup_{u,v\in [0,1]\, : \, |u-v|\le \delta_m} |\omega_v - \omega_u| \ge 2\sqrt{\delta_m\log \delta_m^{-1}} \Bigr\} \le \frac1m .
\end{align*}
Thus given any $\theta >0$, for all $m$ sufficiently large that
$2\sqrt{\delta_m \log \delta_m^{-1}}\le \theta$,
\begin{align*}
	&\limsup_{n\to\infty} \P\left\{ \sup_{u,v\in [0,1]\, : \, |u-v|\le G_n} \left| W_v^{(n)} - W_u^{(n)} \right| \ge \theta \right\}\\
	&\le \limsup_{n\to\infty} \Bigl[ \P\left\{ G_n \ge \delta_m \right\} \, + \,
	\P\Bigl\{ \sup_{\begin{smallmatrix}
			u,v\in [0,1]: \\|u-v|\le \delta_m
	\end{smallmatrix}} \left| W_v^{(n)} - W_u^{(n)} \right| \ge 2 \sqrt{\delta_m \log \delta_m^{-1}} \Bigr\} \Bigr]\\
	&=\P\Bigl\{ \sup_{\begin{smallmatrix}
			u,v\in [0,1]: \\|u-v|\le \delta_m
		\end{smallmatrix}} \left| \omega_v - \omega_u \right| \ge 2 \sqrt{\delta_m \log \delta_m^{-1}} \Bigr\}\\
	&\le \frac{1}{m} .
\end{align*}
Since $m$ is arbitrary it follows that the limit is 0, and so we have shown that $\sup_{u\in \eu{A}_n} W_u^{(n)}$ converges in distribution to $\max_{u\in [0,1]} \omega_u$. By the reflection
principle this is the same as the distribution of $|\omega_1|$, hence has expectation
$\sqrt{2/\pi}$, completing the proof of \eqref{E:Wmax}.

By an application of Doob's Maximal Inequality \cite[Theorem 5.4.3]{rD10}
$$
	\E\bigl[ \max_{u\in [0,1]} W_u^2 \bigr] \le 4  \E [ W_1^2]=4.
$$

\end{proof}

\newcommand{\tZ}{\widetilde{Z}}
\section{Proof of the upper bound}  \label{sec:equalupper}
We replace $A_{t}(i,j)$ by $\max_{i',j'}A_{t}(i',j')\vee 1$ for all $i\ne j$. 
This can only increase the value of $a(\epsilon)$,
so it suffices to prove the upper bound under this new condition. Similarly,
the upper bound will only be increased if we add $\mu - \E[X_t^{(j)}]$ to each
$X_{t}^{(j)}$, so it will suffice to prove the upper bound under the assumption
that the expectations are all the same. 

Throughout this section we use the machinery of Orlicz norms and chaining.
The details of this method are described in section \ref{sec:Orlicz}, but we summarize here the relevant notation.
We write $\|\cdot\|_\Psi$ for the Orlicz norm with respect to the convex function $\Psi=\Psi_2:x\mapsto \e^{x^2}-1$ (so that its inverse is $\Psi^{-1}(y)=\sqrt{\log(1+y)}$).
We equip $\F_T$ with a metric 
\begin{equation}
	\label{E:definerho}
	\rho(f,f'):=\tau\|f-f'\|_2,
\end{equation}
where $\|f-f'\|_2$ is the square root of the Hamming distance between $f$ and $f'$ (as in section \ref{sec:Trajectories}).
When we refer to packing numbers these will be understood to be with respect to this metric.
We define a constant $\tau$ such that $\tau^2/6$ is an upper bound on the {\em variance factor} of $X_t^{(i)} - X_t^{(j)}$ (also defined in section \ref{sec:Orlicz}); 
we could take $\tau^2=24 \max_j \tau_j^2$, where $\tau_j^2$ is a variance factor for $\tX_t^{(j)}$.

Define $\tZ_f:= \sum_{t=0}^{T-1} (X_t^{(f_t)} - X_t^{(0)})$, the main term in $Z_f$. 
Then
$$
\tZ_{f} - \tZ_{f'}=\sum_{t=0}^{T-1} 
\bigl( X^{(f_{t})}- X^{(f'_{t})}\bigr) 
$$
and by Lemma \ref{L:orlicz}
$$
\| \tZ_{f}- \tZ_{f'}\|_{\Psi} \le \rho(f,f').
$$

We now fix a sequence of integers ${1=m_{0}<m_{1}<\cdots<m_{J}<m_{J+1}=T+1}$,
to be determined later, where we assume that $m_{J}=\lfloor T/2\rfloor$ and $m_1 \ge \log T$.
Write ${u_j:= \sqrt{m_j/T} \in (0,1]}$.
We define for $J\ge j\ge 0$,
\begin{equation}  \label{E:alphadef}
\begin{split}
Z_{*}^{j}&:=\max\{ Z_f \, : \, f\in \bigcup_{m_{j}\le k< m_{j+1}} \F_{T,k} \},\\
\tZ_{*}^{j}&:=\max\{ \tZ_f \, : \, f\in \bigcup_{m_{j}\le k< m_{j+1}} \F_{T,k} \}.
\end{split}
\end{equation}
We then have 
\begin{equation} \label{E:ZandZtilde}
	Z_*^{j} \le \tZ_*^{j} - m_j \log \epsilon^{-1} + 2m_{j+1} A^* .
\end{equation}

We may then use \eqref{E:allterms2} to obtain
\begin{equation}  \label{E:Ynbound1}
\begin{split}
a(\epsilon)-a(0) \le 
  &\limsup_{T\to\infty} T^{-1}\log \biggl( 1+\\
  &\quad \sum_{j=0}^{J-1} \epsilon^{m_{j}} (m_{j+1}-m_{j})\binom{T}{m_{j+1}} \e^{2m_{j+1} A^* + \tZ^*_{j}} \\ &\hspace*{3cm}+T\epsilon^{m_{J}} \binom{T}{m_{J}} \e^{2T A^* + \tZ^*_{J}} \biggr) .
\end{split}
\end{equation}

To bound the Orlicz norm of $Z_{*}^{j}$ we use {\em chaining}, as described in
\cite{dP90}. By Lemma \ref{L:chaining} we know that for any $\F_{*}\subset \F_T$,
since $\tZ_{f^{(0)}}$ is identically 0,
\begin{equation}  \label{E:Zstarorlicz}
	\begin{split}
		\|\max_{f\in\F_{*}}\tZ_{f}\|_{\Psi} &\le c_\Psi \int_0^{\diam(\F_*\cup \{f^{(0)}\})} 
		\sqrt{ \log \left(1+ D(r ; \F_*\cup \{f^{(0)}\}) \right) } \dif r \\
		&\le c_\Psi \int_0^{\tau\sqrt{T}} 
		\sqrt{ \log \left(2+ D(r ; \F_*) \right) } \dif r 
	\end{split}
\end{equation}
since the distance between two trajectories is no more than $\tau$ times the square root of the maximum Hamming distance $T$, and adding a single point can increase the packing number by at most 1.

The packing numbers for a union may be bounded by the sum of the individual packing numbers, so we may bound the packing number of $\F_*$ by the sum of the packing numbers for $\F_{T,k}$.
While these are difficult to calculate precisely, particularly for large $k$, in this case we can make do with fairly crude bounds, such as we state below as Lemma \ref{L:packingbound}. 
As the bounds in \eqref{E:packingbound} are increasing in $k$ (for $k\le T$), the sums over an interval such as $\{m_j,m_j+1,\dots, m_{j+1}-1\}$ may be bounded by $m_{j+1}-m_j$ times the bound at $k=m_{j+1}-1$; and then, {\em a fortiori}, by replacing $m_{j+1}-m_j$ by $T$.
We see then that for $j\le J-1$, for any $u>0$
$$
D\Biggl( u\tau\sqrt{T} \; ;\,\bigcup_{m_{j}\le  k < m_{j+1}}\F_{T,k} \Biggr)\le 
T d^{m_{j+1}}\min\left\{ \frac{\e}{u_{j+1}^2}, \frac{\e}{u^2-u_{j+1}^2}+2\e \right\}^{m_{j+1}-1}
$$
and
$$
D\Biggl( u\tau \sqrt{T} \; ;\, \bigcup_{m_{J}\le k\le T}\F_{T,k} \Biggr)\le 
dT (2 d )^{T}
$$

For $j\le J-1$, using the crude bounds $\sqrt{a+b}\le \sqrt{a}+\sqrt{b}$ and $\log(a+b)\le 1+ \log a + \log b$ for all $a,b\ge 1$,
\begin{align*}
\|\tZ_{*}^{j}\|_{\Psi}&\le c_\Psi \tau \sqrt{T} \left( 1 + \sqrt{\log T} + \sqrt{m_{j+1} \log d} \right) \\
&\qquad + 	c_\Psi \tau \sqrt{T} \biggl( \int_{0}^{u_{j+1}} \log^{1/2}\left( \frac{\e}{u_{j+1}^2} \right)^{m_{j+1}}  \dif u\\
&\qquad\qquad  +  \int_{u_{j+1}}^1 \log^{1/2}   \left(\frac{\e }{u^2 - u_{j+1}^2} +2\e \right)^{m_{j+1}} \dif u \biggr)\\
  &\le \sqrt{T m_{j+1}} \left( C_1 + u_{j+1}  \left( 1-  2\log u_{j+1} \right)^{1/2} + \int_0^1 \left( 1- 2\log u \right)^{1/2} \dif u \right)\\
&\le C_{2}\sqrt{Tm_{j+1}}
\end{align*}
for some constants $C_{1},C_{2}$ depending only on $\tau$ and $d$.
By choosing $C_{2}$ appropriately we may ensure that this bound holds as well for $j=J$.

By definition of the Orlicz norm, stated as \eqref{E:orlicz}, this means that for $1\le j\le J$,
$$
\E\left[\exp\left\{(\tZ_{*}^{j})^{2}/C_{2}^{2}Tm_{j+1}\right\}\right] \le 2.
$$
Applying Markov's inequality we have
$$
\P\{ \tZ_{*}^{j} > z\sqrt{Tm_{j+1}}\}\le 2\e^{-z^{2}/C_{2}^{2}}.
$$
Let $z=\max\{1,1.05\cdot C_{2}\}$, and for any $T\ge \log^{2}\epsilon^{-1}$ define $A_{z,T}$ to be the event comprised of
outcomes for which ${\tZ_{*}^{j}\le z\sqrt{Tm_{j+1}(j+1)}}$ for all $j$. Note that
\begin{equation} \label{E:Azprob}
\P(A_{z,T}^{\complement})\le \sum_{j=1}^{\infty} 2\e^{-z^{2}j/C_{2}^{2}}\le 2\left(\e^{z^{2}/C_{2}^{2}}-1\right)^{-1}.
\end{equation}
This bound is smaller than 1, from which it follows that $\P(A_{z,T})>0$.

We now take $m_{j}:=\lfloor 4Tj z^{2}/ \log^{2}\epsilon \rfloor$ as long as this is $<T/2$,
then set $m_{J}=\lfloor T/2\rfloor$ and $m_{J+1}=T+1$. We note that $J\le ( \log^{2}\epsilon)/8z^{2} + 1 $.
By the constraint on $T$, and since $jz^2\ge 1$, we have for $J-1\ge j\ge 1$
$$
\frac{T\e}{m_{j}}\le \frac{\log^{2}\epsilon}{jz^{2}} \cdot \frac{\e}{4 -\frac{\log^{2}\epsilon}{T}}\le
  \frac{ \log^{2} \epsilon}{j z^{2}},
$$
so
$$
\binom{T}{m_{j}}\le \left(  \frac{ \log^{2} \epsilon}{j z^{2}} \right)^{4Tj z^{2}/\log^{2}\epsilon}
  \le \exp \left\{ 8Tj z^{2}\frac{\log\log \epsilon^{-1}}{\log^{2}\epsilon} \right\}
$$
for $1\le j \le J-1$, and 
$$
\binom{T}{m_J}=\binom{T}{\lfloor T/2\rfloor } < 2^{T/2}.
$$
Similarly, we have on $A_{z,T}$ the bounds
\begin{align*}
\tZ_{*}^{j} &\le z\sqrt{Tm_{j+1}(j+1)}\le 2z^{2}(j+1)T/\log\epsilon^{-1} \text{ for } 1\le j \le J-1 \text{ and}\\
\tZ_*^{J} & \le \frac{T \log \epsilon^{-1}}{\sqrt{8}} + (1\vee 2C_2) T.
\end{align*}
Substituting into \eqref{E:Ynbound1} we see that on the event $A_{z,T}$,
\begin{align*}
&a(\epsilon)-a(0)\\
&\le \liminf_{T\to\infty}T^{-1}\log \biggl( 1 +T \exp\biggl\{ \frac{T}{2} \Bigl( \log 2 +4 A^* +(2\vee 4C_2) - ( 1 -\frac{1}{\sqrt{2} })   \log \epsilon^{-1}\Bigr)\biggr\} \\
&\quad +T\sum_{j=1}^{\infty} 
	\exp \biggl\{ -\frac{2T z^2}{\log\epsilon^{-1}} \Bigl( 2(j-1) - 4j \frac{\log\log \epsilon^{-1}}{\log \epsilon^{-1}} - \frac{4A^*  j }{\log\epsilon^{-1}} - j  \Bigr)  \biggr\}  \biggr).
\end{align*}
We restrict now to $\epsilon$ sufficiently small that
\begin{equation} \label{E:epsilon_small}
\frac{\log \epsilon^{-1}}{A^*+ \log \log \epsilon^{-1}} > 8 \quad \text{and} \quad   \log \epsilon^{-1} > \frac{\log 2 +4 A^* +(2\vee 4C_2)}{1-\frac{1}{\sqrt{2}}}
\end{equation}
The sum may then be bounded by
$$
\sum_{j=-3}^\infty \exp\biggl\{ - \frac{Tz^{2}}{\log\epsilon^{-1}} j \biggr\}
  = \frac{ \exp\left\{ 4 Tz^{2}/\log\epsilon^{-1}  \right\} }{ \exp\left\{ Tz^{2}/\log\epsilon^{-1}  \right\} -1},
$$
while the terms on the first line are bounded by $1+ T \e^{-T/2} \le 2$
for $\epsilon$ in the stated range.
Thus
\begin{equation}  \label{E:Ynbound3}
\begin{split}
a(\epsilon)-a(0)
&\le \liminf_{T\to\infty}T^{-1}
  \log \left( 2 + \frac{\log \epsilon^{-1}}{T z^2} \exp\biggl\{ \frac{Tz^{2}}{\log\epsilon^{-1}}  \biggr\}\right)\\
  &=\frac{z^2}{\log\epsilon^{-1}}
\end{split}
\end{equation}
on the event $A_{z,T}$. Since the event has positive probability, and since
$a(\epsilon)-a(0)$ is almost surely constant, the bound holds with probability 1.

\begin{lemma} \label{L:packingbound}
For positive integers $T,k$, and any positive $r$ with $\tau\sqrt{T} \ge r > \tau \sqrt{k}$,
\begin{equation}  \label{E:packingbound}
D(r \, ;\, \F_{T,k})\le d^{k+1} \left( \frac{T\e}{(r/\tau)^{2}-k}+2\e \right)^{k}.
\end{equation}
and for all $r$
\begin{equation} \label{E:packingbound2}
D(r \, ;\, \F_{T,k}) \le d^{k+1}\left( \frac{T\e}{k} \right)^k
\end{equation}
\end{lemma}

\begin{proof}
Let $r'=(r/\tau)^{2}$. Suppose that $r'>k$, let $j=\lfloor r'/k\rfloor$, and let $m=\lceil T/j\rceil$. Let $\F_{*}$ be the set of ordered (non-decreasing) sequences
of length $k$ from $\{0,\dots,m-1\}$, crossed with $\{0,\dots,d-1\}^{k+1}$, 
and define a map $(\phi,\psi):\F_{T,k}\to \F_{*}$ 
by letting $\{f\}_{i}$ be the $i$-th coordinate where $f$ changes, and defining
$$
\phi(f)_{i}= \bigl\lfloor \frac{\{f\}_{i}}{j}  \bigr\rfloor
$$
and $\psi(f)_{i}=f_{\{f\}_{i}}$; that is, the site that $f$ moves to at its $i$-th
change.

If $f$ and $f'$ are two elements of $\F_{T,k}$ with $\phi(f)=\phi(f')$
and $\psi(f)=\psi(f')$, 
then $f_{i}=f'_{i}$ as long as $\lfloor i/j\rfloor\notin \phi(f)$, 
since any $t\notin \phi(f)$ corresponds to a span of $tj,tj+1,\dots,tj+j-1$
where $f_{tj}=f'_{tj}$ (because they started with $f_{0}=f'_{0}$, and the number
of changes in $f_{0},\dots,f_{tj-1}$ is the same as the number
of changes in $f'_{0},\dots,f'_{ti-1}$). 
Thus the Hamming distance is no more than $k\cdot j \le r'$, meaning that $\|f-f'\|_{2}\le\sqrt{r'}$, so $\rho(f,f')\le \tau\sqrt{r'}=r$. By the
pigeonhole principle, any subset of $\F_{T,k}$ of size greater than $\# \F_{*}$
has points with $\rho$-separation no more than $r$. Hence
$$
D(r,\F_{T,k})\le \#\F_{*} \le \binom{m+k}{k} d^{k+1}.
$$

Combining this with the trivial bound $D(r,\F_{T,k})\le \#\F_{T,k}\le d^{k+1}\binom{T}{k}$ and the bound
$$
\binom{a}{b}\le \left(\frac{a\e}{b}\right)^{b}
$$
completes the proof.
\end{proof}

\section{Simulations}  \label{sec:simulation}
We illustrate the result with a very simple $2\times 2$ example:
\begin{align*}
&\Les_{t}=
\begin{pmatrix}
 0& B_{t}\\
 S_{t}&0
\end{pmatrix}, \qquad
A_{t}=
\begin{pmatrix}
1&0\\0&0
\end{pmatrix}, \\
&\text{with }
S_{t}\sim \operatorname{Unif}(0.05,0.99),\quad B_{t} \sim \operatorname{Gam}(5,2),
\end{align*}
with $S_{t}$ and $B_{t}$ independent. We have $a(0)=\frac12(\E[\log S_{t}]+\E[\log B_{t}])=-0.0193$. We also have
$$
	\sigma_*^2 = \var(\log S_t/B_t) + \var(\log B_t/S_t)=2\var(\log S_t)+2\var(\log B_t) \approx 1.444.
$$ 

We calculate average log growth rates over simulations of $10^6$ timesteps.
With 1000 replicates the standard errors for the $a(\epsilon)$ estimates are all below $2\times 10^{-5}$, and substantially less than 1\% of the mean estimates, hence the 4 digits shown here may be considered reliable.

\begin{table}[ht]
\centering
\begin{tabular}{ccc}
  \toprule
 $\epsilon$&$1/\log\epsilon^{-1}$&$a(\epsilon)$\\
  \midrule
0.5 & 1.4427 & 0.3052 \\ 
0.4 & 1.0914 & 0.2561 \\ 
0.3 & 0.8306 & 0.2057 \\ 
0.2 & 0.6213 & 0.1532 \\ 
0.1 & 0.4343 & 0.0968 \\ 
0.05 & 0.3338 & 0.0646 \\ 
0.01 & 0.2171 & 0.0296 \\ 
0.005 & 0.1887 & 0.0220 \\ 
0.001 & 0.1448 & 0.0109 \\ 
$10^{-4}$ & 0.1086 & 0.0026 \\ 
$10^{-5}$ & 0.0869 & -0.0022 \\ 
$10^{-6}$ & 0.0724 & -0.0052 \\ 
$10^{-7}$ & 0.0620 & -0.0073 \\ 
$10^{-8}$ & 0.0543 & -0.0089 \\ 
$10^{-9}$ & 0.0483 & -0.0101 \\ 
$10^{-10}$ & 0.0434 & -0.0111 \\ 
0 & 0.0000 & -0.0192 \\ 
   \bottomrule
\end{tabular}
\caption{Simulated $2\times 2$ average log growth rates based on 1000 simulations of $10^6$ timesteps for each $\epsilon$.
Matrices are determined by i.i.d.\ samples from $S_{t}\sim \operatorname{Unif}(0.05,0.99),\, B_{t} \sim \operatorname{Gam}(5,2)$.} 
\label{T:diapausesim}
\end{table}

\begin{figure}[h]
\begin{center}
\includegraphics[width=11cm]{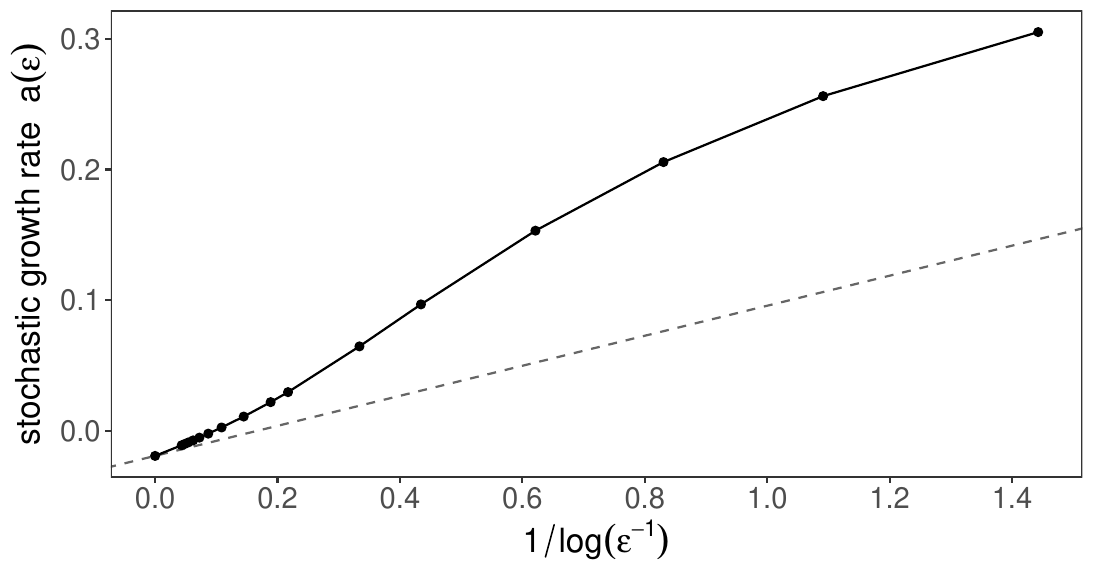}
\caption{Plot of results from Table \ref{T:diapausesim}.
Dashed line is the asymptotic lower bound, the line through $(0,a(0))$ with slope $0.115$.}
\label{F:diapausesim}
\end{center}
\end{figure}

The results are tabulated in Table \ref{T:diapausesim}, for values of $\epsilon$ down to $10^{-10}$. In Figure \ref{F:diapausesim} we plot $a(\epsilon)$
against $1/\log \epsilon^{-1}$, and see that for small values of $\epsilon$ the values are very close to a line, with slope approximately $0.2$. This is consistent with Theorem \ref{T:samerate1}, which states that it should
converge (as $1/\log\epsilon^{-1}\to 0$) to a line with
slope at least $\sigma_*^2/2d\pi = 0.115$.
Empirically, we find that the slope from $\epsilon=0$ to $\epsilon=10^{-10}$ is $(-0.0111-(-0.0192))/(0.0434-0)=0.188$, so the lower bound appears to hold appropriately, and to be of approximately the right order in this case.

\appendix
\section{Sub-Gaussian random variables, the Orlicz norm, and chaining} \label{sec:Orlicz}
For purposes of the upper bound on the sensitivity of $a(\epsilon)$ we will need to
assume that the growth rate discrepancies $\tX_t^{(j)}:= X_t^{(j)}-X_t^{(0)}$ are sub-Gaussian.
We collect below some facts about application of the method of {\em chaining} to bound maxima of sub-Gaussian random variables.
Recall that a random variable $Z$ is said to be {\em sub-Gaussian} if it
has finite {\em variance factor} (the term applied in \cite{BLM13}) $\tau^2=\tau^2(Z)$, defined by
\begin{equation} \label{E:sGs}
	\tau(Z):=\inf \left\{ c\ge 0\,:\, \log \E\left[\e^{\lambda (Z - \E[Z])}\right] \le 
	\frac{c^2\lambda^{2}}{2} \, \forall \lambda \in\R\right\}.
\end{equation}
($\tau(Z)$ itself
is called the {\em sub-Gaussian standard} in \cite{BK00}.)
This may be thought of as an upper 
bound on the scale of the tails, and it is this that determines the upper bound on the sensitivity of the stochastic growth rate
to migration.
Note that the Legendre transform for the moment generating function
$I_Z(z):= \sup_{\lambda}  \{\lambda z - \log \E[\e^{\lambda (Z-\E[Z])}]\} $ satisfies
$I_Z(z) \ge z^2/2 \tau^2(Z) $ for all $z$. Applying Markov's inequality implies the
standard sub-Gaussian concentration inequality
\begin{equation} \label{E:conc_ineq}
	\P\bigl\{ Z \ge  \E[Z]  + z \bigr\} \le \e^{-z^2/2\tau^2 (Z)} .
\end{equation}
Of course, Gaussian random variables with variance $\tau^2$ are also sub-Gaussian with variance factor $\tau^2$, but the class of sub-Gaussian random variables is much larger.
We write $\tau_j^2$ for the variance factor of $\tX_t^{(j)}$.

Following \cite{VW96} we define the Orlicz norm 
$\|Z\|_{\Psi}$ for any random variable $Z$ and any convex increasing function $\Psi$ by
\begin{equation} \label{E:orlicz}
	\|Z\|_\Psi:=\inf\{ C \,:\, \E[\Psi(|Z|/C)]\le 1\}.
\end{equation}
Most commonly $\Psi$ is taken to be $\Psi_\alpha: x \mapsto \e^{x^{\alpha}} -1$ for some $\alpha \ge 1$. 
The Orlicz norm has useful properties when applied to maxima
of arbitrary collections of random variables. 
Using the method of {\em chaining} this may be extended to suprema of processes
indexed by an arbitrary finite-dimensional semi-metric space $(\F,\rho)$.
The basic idea is that the maximum of a separable stochastic process\footnote{The processes considered in this paper are
	continuous, hence automatically separable.} may be approximated in shrinking increments by the maxima over
increasingly dense ``skeletons'' of $\F$. 
The higher the dimension of the space, the more rapid the growth of the skeletons.
This is described by the {\em packing numbers} $D(r; \F,\rho)$, defined to be the maximum number of points
that may be selected from $\F$, such that for any $f,f'$ among them $\rho(f,f')\ge r$. 
As it will generally be clear from context which metric is meant, we will drop $\rho$ from the notation
and write simply $D(r;\F)$.
We then have the following lemma \cite[Corollary 2.2.5]{VW96}:
\begin{lemma} \label{L:chaining}
	There is a constant $c_\Psi$, depending only on the choice of $\Psi$,
	such that for any separable real-valued stochastic process $(Z_f)_{f\in \F}$ satisfying $\| Z_f - Z_{f'} \|_\Psi \le \rho(f,f')$,
	\begin{equation} \label{E:chaining}
		\bigl\| \sup_{f,f'\in\F} |Z_f - Z_{f'}| \bigr\|_\Psi \le c_\Psi \int_0^{\diam(\F)} \Psi^{-1} \bigl( D(r ; \F, \rho ) \bigr) \dif r \, .
	\end{equation}
\end{lemma}

We now fix $\Psi=\Psi_2: x \mapsto \e^{x^2}-1$. 
A random variable is sub-Gaussian if and only if its Orlicz norm with this choice of $\Psi$ is finite.
The Orlicz norm is a norm, hence sub-additive, so that $\| Z_1 + Z_2\|_\Psi \le \| Z_1\|_\Psi + \| Z_2\|_\Psi$.
When the random variables are independent we have a stronger result
(which is a variation on Lemma 1.7 of \cite{BK00}.)
\begin{lemma} \label{L:orlicz}
	If $Z_1,Z_2,\dots,Z_n$ are independent mean-zero sub-Gaussian random variables
	with variance factors $\tau^2_1,\dots,\tau^2_n$, then
	\begin{equation} \label{E:orlicznbound}
		\left\| Z_1 + \cdots + Z_n\right\|_\Psi
		\le \sqrt{6(\tau^2_1+\cdots + \tau^2_n)} .
	\end{equation}
\end{lemma}

\begin{proof}
	Let $X=Z_1+\cdots + Z_n$ and $\tau^2 = n^{-1} (\tau^2_1 + \cdots + \tau^2_n)$. Then for all real $\lambda$ we have
	$$
	\E\left[ \e^{\lambda X} \right] \le \e^{n\tau^2 \lambda^2/2}
	$$
	hence, taking $\lambda= \pm z/n\tau^2$,
	$$
	\P\left\{ |X| \ge z \right\} \le 2\e^{n\tau^2 (z/n\tau^2)^2/2 - (z/n\tau^2) z} = 2\e^{-z^2/2n\tau^2}
	$$
	for all $z >0$. We have then
	\begin{align*}
		\E\left[ \Psi(X/C) \right] & = \int_0^\infty \frac{1}{C} \Psi'(z/C) \P\left\{ |X| \ge z \right\} \dif z\\
		&\le \int_0^\infty \frac{4z}{C^2} \e^{\frac{z^2}{C^2} - \frac{z^2}{2n\tau^2} }\dif z\\
		& = \frac{4n\tau^2}{C^2 - 2n\tau^2},
	\end{align*}
	which is $\le 1$ for $C\ge \sqrt{6n\tau^2}$.
\end{proof}

\section*{Acknowledgements}
We thank Stanford University and University of Oxford as our funding sources.
For the purpose of Open Access, the authors have applied a CC BY public copyright license to any Author Accepted Manuscript version arising from this submission.

\bibliographystyle{alpha}
\bibliography{diamig}
\end{document}